\documentclass{article}\date{}\begin{document}

\title{Chern-Simons ``ground state" from the path integral}

\author{Z. Haba\\
Institute of Theoretical Physics, University of Wroclaw,\\ 50-204
Wroclaw, Plac Maxa Borna 9, Poland\\
email:zbigniew.haba@uwr.edu.pl}\maketitle
\begin{abstract} We consider a path integral representation
 of the time
evolution $\exp(-\frac{i}{\hbar}tH)$ for Lagrangians of the
variable $A$ which can be represented in the form  (quadratic in
$Q$) ${\cal L}(A)=\frac{1}{2}Q(A){\cal
M}Q(A)+\partial_{\mu}L^{\mu}$. We show that

$\exp(-\frac{i}{\hbar}tH)\exp(\frac{i}{\hbar}\int d{\bf x}L^{0})
=\exp(\frac{i}{\hbar}\int d{\bf x}L^{0})$ up to an $A$-independent
factor. We discuss examples of the states
$\exp(\frac{i}{\hbar}\int d{\bf x}L^{0})$ in quantum mechanics and
in quantum field theory (the Chern-Simons states in Yang-Mills
theory, Kodama states in quantum gravity). We show the relevance
of these states for a determination of the dynamics in terms of
stochastic perturbations of self-duality equations. The solution
of the Schr\"odinger equation can be expressed by the solution of
the self-duality equation in the leading order of $\hbar$
expansion. We discuss applications to gauge theory on a Lorentzian
manifold and gauge theories of gravity.
\end{abstract}

\section{Introduction}
The Feynman path integral can either be derived from Hamiltonian
dynamics or postulated as a quantization principle. In gauge
theories the correct quantization has been derived by a precise
determination (the Faddeev-Popov formula)   of the formal Feynman
measure. It usually can be obtained by means of a proper
extraction of independent variables or by an imposition of
constraints allowing to restrict the  variables to their physical
range \cite{dirac}. In this paper we show that by a transformation
of the Lagrangian to a quadratic form plus an eventual total
divergence term, we can find on the basis of the path integral, a
particular state which is transformed just by a trivial phase
factor under the time evolution. Vice versa this state determines
a transformation of the Lagrangian to new variables which make the
Lagrangian quadratic. In mathematics such transformations are
known in the theory of diffusion processes. In physics they
appeared in supersymmetric quantum mechanics \cite{super} and in
supersymmetric field theories
\cite{parisisourlas}\cite{nicolai}.We have extended the method to
scalar field theories in \cite{habaepj}\cite{habaentropy}.

In this paper we discuss in detail transformations of the path
integral in simple Hamiltonian models (sec.2). We check that the
transformed path integral leads to the correct solution of the
Schr\"odinger equation. Subsequently, we apply the method to the
electromagnetic field on a globally hyperbolic manifold
\cite{global}. Then, the Lagrangian can be expressed as a square
of the (anti) self-dual field strength plus a topological
invariant ( a divergence of the Chern-Simons term). It follows
from the path integral that the resulting Chern-Simons state
\cite{witten} does not change under time evolution. Such states
lead to interesting phenomena in condensed matter physics in 2+1
dimensions related to anyons \cite{wilczek} and charged vortices
with fractional statistics \cite{frohlich}. The Chern-Simons
Lagrangian \cite{witten1}\cite{witten2} is a basis of a
topological field theory with a prospective application to quantum
gravity. In the Abelian gauge theory we show that the self-dual
field strength can be treated as a new Gaussian variable. The
solution of the Schr\"odinger equation can be expressed by a
solution of a stochastic perturbation of the self-duality
equation. From the path integral it follows that the Chern-Simons
wave function is the time-independent solution of the
Schr\"odinger equation for a quantum electromagnetic field on a
manifold. By means of an explicit calculation in the Hamiltonian
formulation we check this conclusion. The method can be extended
to non-Abelian gauge theories. We discuss the Yang-Mills theory
with de Sitter group as an internal symmetry group in view of its
potential application to a quantization of gravity
\cite{mansouri}\cite{townsend}\cite{don}. We show that the
Chern-Simons wave function in this model is a solution of the
Schr\"odinger equation with an arbitrary choice of the metric (in
particular the one  identifying the conformal tetrad with the one
on the base manifold). Then, a solution of the Schr\"odinger
equation with the Chern-Simons wave function as the WKB factor can
be expressed by the solution of the self-duality equation in the
leading order of $\hbar$ expansion. We believe that the change of
variables in the path integral can serve as an efficient tool in
quantization of systems with a large symmetry group. The eventual
constraints on the variables of the model can be imposed on the
stochastic equations restricting the dynamics to the constrained
phase space.

\section{The path integral in simple models}
We consider a  path integral for a system described by a variable
$A$ with  a Lagrangian ${\cal L}$. Then, the time evolution of the
state $\psi(A)$ is given by the formula
\begin{equation}
\psi_{t}(A)=\int {\cal
D}A\exp\Big(\frac{i}{\hbar}\int_{0}^{t}d{\bf x}ds{\cal
L}(A_{s})\Big)\psi(A_{t}(A)).
\end{equation}
 Assume that the Lagrangian  can be expressed in the form
\begin{equation}
L=\int d{\bf x}{\cal L}=\int d{\bf x}(\frac{1}{2}Q(A){\cal
M}Q(A)+\partial_{\mu}L^{\mu}),
\end{equation}where ${\cal M}$ does not depend on $A$.
Then, the path integral (1) reads (we assume that the spatial part
of $L^{\mu}$ is vanishing at infinity)
\begin{equation}
\begin{array}{l}
\psi_{t}(A)=\exp\Big(-\frac{i}{\hbar}\int d{\bf
x}L^{0}(A)\Big)\int {\cal
D}A\exp\Big(\frac{i}{\hbar}\int_{0}^{t}d{\bf
x}ds\frac{1}{2}Q(A){\cal
M}Q(A)\Big)\cr\times\exp\Big(\frac{i}{\hbar}\int d{\bf
x}L^{0}(A_{t}(A))\Big)\psi(A_{t}(A)).\end{array}
\end{equation}
We denote \begin{displaymath} W=i\int d{\bf x}L^{0}(A).
\end{displaymath}
We express the initial state $\psi$ in the form
\begin{equation}
\psi=\exp(-\frac{1}{\hbar}W)\chi.
\end{equation}  Then\begin{equation}\begin{array}{l}
\psi_{t}(A)=\exp(-\frac{W}{\hbar})\int {\cal
D}A\exp\Big(\frac{i}{\hbar}\int_{0}^{t}d{\bf
x}ds\frac{1}{2}Q(A){\cal M}Q(A)\Big)\chi(A_{t}(A))\equiv
\exp(-\frac{W}{\hbar})\chi_{t}(A).\end{array}
\end{equation}
On a formal level, if the Jacobian $\frac{\partial Q}{\partial
A}\simeq const$, then we can change variables $ A \rightarrow
Q(A)$ so that
\begin{displaymath}
\chi_{t}(A)=\Big< \chi (A(t,Q))\Big>,
\end{displaymath}
where $Q$ is a Gaussian variable. We explain the formal procedure
in the precise way in simple models below.

If we choose $\chi=1$ then $\psi=\exp(-\frac{1}{\hbar}W)$ and
eq.(5) reads
\begin{equation}
\begin{array}{l}
\psi_{t}(A)=\psi(A)\int {\cal
D}A\exp\Big(\frac{i}{\hbar}\int_{0}^{t}d{\bf
x}ds\frac{1}{2}Q(A){\cal M}Q(A)\Big).\end{array}
\end{equation}
Hence, $\psi_{t}(A)=Z_{t}\psi(A)$, where the factor $Z_{t}$ does
not depend on $A$. By the unitary evolution law
$U_{t}U_{s}=U_{t+s}$ we can conclude that
$Z_{t}=\exp(-\frac{iEt}{\hbar})$ with a real constant $E$. If the
Jacobian $\frac{\partial Q}{\partial A}\simeq const$ (as will be
the case in our examples), then the integral on the rhs of eq.(6)
is a time independent constant $\det{\cal M}^{-\frac{1}{2}}$.
 In such a case it
follows from eq.(6) that $\psi=\exp(-\frac{1}{\hbar}W)$ does not
change in time. Hence, it is the ground state. We do not assume
that the state $\psi=\exp(-\frac{1}{\hbar}W)$ is square
integrable. In general, it is not. From the change of variables
(2) $A\rightarrow Q$  the quantum Hamiltonian dynamics is
expressed by an equivalent stochastic dynamics which we derive in
simple models in this and in the subsequent sections. In  complex
models with constraint variables (as in various versions of
gravity) the Hamiltonian dynamics may be hard to determine
explicitly  although the time independence of some states (as the
Kodama state \cite{witten}\cite{kodama}) may follow from the path
integral. Then, the constraints can be imposed on the resulting
stochastic dynamics.

The result that $H\exp(-\frac{1}{\hbar}W)=0$ will be confirmed in
several models in this paper. As a first  example let
\begin{equation}\begin{array}{l}
{\cal L}=\frac{1}{2}(\partial_{t}x_{a})^{2}-V(x)
=\frac{1}{2}\Big(\partial_{t}x_{a}+i\partial_{a}W\Big)^{2}-i\frac{dx_{a}}{dt}\partial_{a}W.
\end{array}\end{equation} We assume that $\triangle W=0$ (we
explain in a moment this assumption). Then,
 we can check by elementary calculations that the Lagrangian is of the form
(2) and $H\exp(-\frac{W}{\hbar})=0$ if
\begin{equation}
\frac{1}{2}\partial_{a}W\partial_{a}W=V
\end{equation} and $\triangle W=0$.
The mysterious  (from the point of view of the path integral)
condition of $W$ being a harmonic function
  can be explained either from the
 requirement that  $\int \partial_{a}W dx_{a}
 =\int \partial_{a}W\circ dx_{a}$ (the Ito integral $dx$ is equal to the Stratonovitch integral $\circ dx$
 \cite{ikeda}\cite{simon},then the integral (7) in the action does not depend on the form of
 the discretization) or by a
  computation of the Jacobian $\frac{\partial Q}{\partial
  x}\simeq \exp(\int\triangle W)=const$. The heuristic argument leading to
  the constant Jacobian comes from the computation
\begin{displaymath}\begin{array}{l}
\det(\frac{\partial Q}{\partial
x})=\det\Big(\delta_{ab}\partial_{t}-i\partial_{a}\partial_{b}W)
=\exp\Big(Tr\ln\Big(\delta_{ab}\partial_{t}-i\partial_{a}\partial_{b}W\Big)\Big).\end{array}\end{displaymath}
We expand $Tr\ln
(\big(\delta_{ab}-i\partial_{t}^{-1}\partial_{a}\partial_{b}W\big)$
in powers of $W$  with
$\partial_{t}^{-1}(t,t^{\prime})=\theta(t-t^{\prime})$ (where
$\theta $ is the Heaviside function) and check that in the
expansion the products of $\theta$ functions give a non-zero
result only in the lowest order which is $\int_{0}^{t} ds\triangle
W$ . The model (7) can be described in terms of fermionic degrees
of freedom if $\triangle W\neq 0$.
  Then, it is invariant under supersymmetry transformations
   \cite{super}.

The transformation $A\rightarrow Q$ mentioned at eq.(5) is related
to a  diffusion process. In eq.(7) we have got explicitly the
quadratic representation of the Lagrangian. In fact, if
$H\exp(-\frac{1}{\hbar}W)=0$ , $\psi=\exp(-\frac{1}{\hbar}W)\chi$
and $i\hbar\partial_{t}\psi_{t}=H\psi_{t}$ then $\chi_{t}$
satisfies the equation
\begin{equation}
\partial_{t}\chi_{t}=(\frac{i\hbar}{2}\triangle-i\partial_{a}\ln
W\partial_{a})\chi_{t}
\end{equation}
with the solution
\begin{equation}
\chi_{t}({\bf x})=E[\chi({\bf q}_{t}({\bf x}))],
\end{equation}
where ${\bf q}_{t}({\bf x})$ is the solution of the stochastic
equation (for complex stochastic processes in quantum mechanics
see \cite{hababook})
\begin{equation}
dq_{a}=-i\partial_{a}Wdt+\sqrt{i\hbar}db_{a}
\end{equation}with the initial condition ${\bf x}$. Here, $b_{a} $ is the Brownian motion, i.e.,
the Gaussian process with mean zero and  the covariance (defined
for $t,s\geq 0$)
\begin{equation}E[b_{a}(s)b_{c}(t)]=\delta_{ac}min(t,s).
\end{equation}
We could take as $W$ a real part of a (pluri) holomorphic
function. So, in two-dimensions ($z=x_{1}+ix_{2}$)let the
potential be $V=\frac{9}{2}\lambda^{2}(x_{1}^{2}+x_{2}^{2})^{2}$
then $H\exp(-\frac{W}{\hbar})=0$  if
$W=\frac{\lambda}{2}(z^{3}+\overline{z}^{3})=\lambda(x_{1}^{3}-3x_{1}x_{2}^{2})$.
$\exp(-\frac{W}{\hbar})$ is not square integrable. In order to
have a square integrable solution we must consider initial
conditions $\chi$ decaying sufficiently fast at infinity .

As  another example consider\begin{equation}
W=\frac{\lambda}{2}(z^{4}+\overline{z}^{4})=\lambda\Big(
(x_{1}^{2} +x_{2}^{2})^{2}-8x_{1}^{2} x_{2}^{2}\Big).
\end{equation} In this model $\exp(-\frac{1}{\hbar}W)$ is
decaying fast except of the region $x_{1}\sim x_{2}$. From eq.(8)
the potential is
\begin{displaymath}
V=8\lambda^{2}(x_{1}^{6}+x_{2}^{6})+24\lambda^{2}(x_{1}^{2}+x_{2}^{2})x_{1}^{2}x_{2}^{2}
\end{displaymath}
We can express the solution either by means of the stochastic
process (10)-(11) or by means of the Feynman-Kac formula
\cite{simon}\cite{hababook}
\begin{equation}
\psi_{t}({\bf x})=E\Big[\exp\Big(-\frac{i}{\hbar}\int_{0}^{t}ds
V({\bf x}+\sqrt{i\hbar}{\bf b}_{s})\Big)\psi({\bf
x}+\sqrt{i\hbar}{\bf b}_{t})\Big].
\end{equation}
The exponential in this formula is bounded (hence the expectation
value is finite) because $-\frac{i}{\hbar}V\simeq
-\lambda^{2}\hbar^{2}({\bf b}^{2})^{3}$ for large $\vert{\bf
b}\vert$. This example shows that the models with the initial
condition $\psi=\exp(\frac{-W}{\hbar})\chi$ where $W$ is unbounded
from below can have well-defined stochastic solution. Comparing
eq.(14) with eqs.(5),(10)-(11) we can see that the stochastic
version can be considered as a complex WKB method leading to an
expansion in $\sqrt{\hbar}$ with the expansion coefficients
computable as expectation values over the Brownian motion.

 In field theory in one spatial dimension let us consider
 \begin{equation}\begin{array}{l}
 \int d{\bf x}{\cal L}=\int d{\bf x}(\frac{1}{2}\partial_{t}\phi_{a} \partial_{t}\phi_{a}
 - \frac{1}{2}\partial_{x}\phi_{a} \partial_{x}\phi_{a}
 -V(\phi)) =\int d{\bf x}\Big(\frac{1}{2}(\partial_{t}\phi_{a}
 \cr- i\epsilon^{ab}\partial_{x}\phi_{b} -i\partial_{a}W)^{2}
 +i\partial_{t}W
+\frac{i}{2}\epsilon^{ab}\epsilon^{\mu\nu}\partial_{\mu}(\phi_{a}\partial_{\nu}\phi_{b}))\Big),
 \end{array}\end{equation}where $a,b=1,2$ and we assumed that fields vanish at spatial infinity. $W$ is related to $V$ by eq.(8), $\epsilon$ is the Levi-Civita
 antisymmetric tensor in two-dimensions and $W=iL^{0}=\Re f(\phi_{1}+i\phi_{2})$, where $f$ is a holomorphic
  function of eq.(13).
  The last term in eq.(15) is
 independent of the metric (a topological invariant ). We shall
 encounter such terms in gauge theories in the next section.

As an example of the potential    we consider again
$V=\frac{9}{2}\lambda^{2}(\phi_{1}^{2}+\phi_{2}^{2})^{2}$. We
define $\tilde{W}$ as
\begin{equation}\tilde{W} =\int dx\Big(\frac{\lambda}{2}((\phi_{1}+i\phi_{2})^{3}+(\phi_{1}-i\phi_{2})^{3})
+\phi_{1}\frac{\partial}{\partial x}\phi_{2}\Big).
\end{equation} The formulas (2) and (15) follow from $\partial_{t}\int
d{\bf x}\tilde{W}= \int d{\bf x}\frac{\delta\tilde{W}}{\delta
\phi_{a}}\partial_{t}\phi_{a}$ and
\begin{displaymath}
\int
 dx(V(\phi)+\frac{1}{2}\partial_{x}\phi_{a}\partial_{x}\phi_{a})=\frac{1}{2}\int
dx\frac{\delta \tilde{W}}{\delta\phi_{a}}\frac{\delta
\tilde{W}}{\delta\phi_{a}}.
\end{displaymath}It can be checked that
$H\exp(-\frac{\tilde{W}}{\hbar})=0$.

 For the stochastic
representation (10) in quantum field theory
\begin{displaymath}
\chi_{t}(\phi)=E[\chi(\phi_{t}(\phi))]
\end{displaymath}
we solve the stochastic equations (discussed by Parisi and Sourlas
\cite{parisisourlas} in the Euclidean space)
 \begin{equation}
\partial_{t}\phi_{1}=i\partial_{x}\phi_{2}-i3\lambda(\phi_{1}^{2}-\phi_{2}^{2})+\sqrt{i\hbar}\partial_{t}B_{1},
\end{equation}
\begin{equation}
\partial_{t}\phi_{2}=-i\partial_{x}\phi_{1}+6i\lambda\phi_{1}\phi_{2}+\sqrt{i\hbar}\partial_{t}B_{2},
\end{equation}
These equations are analogs of eq.(11) describing the change of
variables $ \phi\rightarrow Q$ (with $\frac{\partial W}{\partial
q_{a}}\rightarrow \frac{\delta \tilde{W}}{\delta\phi_{a}}$) and
the Gaussian variables $Q_{a}$ identified as  the Brownian motion
$B_{a}$ in two-dimensional space-time which is defined by the
covariance
\begin{displaymath}
E[B_{a}(t,x)B_{b}(s,y)]=min(t,s)\delta_{ab}\delta(x-y).
\end{displaymath}
 Differentiating
eqs.(17)-(18) we obtain a stochastic wave equation
\begin{displaymath}
(\partial_{t}^{2}-\partial_{x}^{2})\phi_{a}+\partial_{a}V(\phi)=\sqrt{\hbar}\eta_{a},
\end{displaymath}
where $\eta_{a}$ is a complex noise.

The total derivative in eq.(2) does not change Lagrange equations
of motion but it does change the canonical formalism (see
\cite{habaepjplus}). The Hamiltonian  modified by the total
derivative is the one on the rhs of eq.(9).
\section{Quantization of gauge theories on a Lorentzian manifold}
We consider the Lagrangian of the electromagnetic field
\begin{equation}\begin{array}{l}
{\cal
L}=-\frac{1}{4}\sqrt{-g}g^{\mu\nu}g^{\alpha\beta}F_{\mu\alpha}F_{\nu\beta}
=-\frac{1}{8}\sqrt{-g}g^{\mu\nu}g^{\alpha\beta}(F_{\mu\alpha}\pm
\frac{i}{2}\tilde{\epsilon}_{\mu\alpha\sigma\rho}F^{\sigma\rho})(F_{\nu\beta}\pm
\frac{i}{2}\tilde{\epsilon}_{\nu\beta\sigma\rho}F^{\sigma\rho})\cr\mp
\frac{i}{4}\epsilon_{\mu\alpha\sigma\rho}F^{\mu\alpha}F^{\sigma\rho}
=-\frac{1}{8}\sqrt{-g}F^{\pm}_{\mu\alpha}F^{\pm\mu\alpha}\mp\frac{i}{4}F^{*}_{\mu\alpha}F^{\mu\alpha}
\end{array}
\end{equation}
defined on a Lorentzian manifold with the metric $g_{\mu\nu}$ ($g$
denotes the determinant of the metric). In eq.(19)
\begin{equation}
\tilde{\epsilon}_{\mu\alpha\sigma\rho}=\sqrt{-g}\epsilon_{\mu\alpha\sigma\rho},\end{equation}
where on the lhs we have the tensor density $\tilde{\epsilon}$ and
on the rhs the Levi-Civita antisymmetric symbol on the Minkowski
space-time defined by $\epsilon^{0123}=1$,
$F^{*}_{\mu\nu}=\epsilon_{\mu\nu\alpha\beta}F^{\alpha\beta}$.

The representation of the Lagrangian (19) is of the form (2) as
\begin{equation}
\frac{i}{2}\epsilon_{\mu\alpha\sigma\rho}F^{\mu\alpha}F^{\sigma\rho}=i\partial_{\mu}
(A^{\alpha}\epsilon_{\mu\alpha\sigma\rho}F^{\sigma\rho})\equiv
i\partial_{\mu}L^{\mu}.
\end{equation}
As a consequence of eq.(5) the Chern-Simons state
\begin{equation}
\psi_{CS}=\exp(\pm\frac{1}{2\hbar}\int d{\bf
x}A_{j}\epsilon^{jkl}\partial_{k}A_{l})
\end{equation}
is an eigenstate. In fact, it does not change in time because the
Jacobian $\frac{\partial Q}{\partial A}\simeq \exp\int
\frac{\partial Q}{\partial A\partial A}=const$ ( the argument
follows the one leading to eq.(11) as $Tr \frac{Q_{jk}}{\partial
A_{n}\partial A_{n}}=0 $). Let us note that
$\partial_{\mu}L^{\mu}$ does not depend on the metric (it is a
topological invariant; the Chern-Simons term).

We check directly that $\psi_{CS}$ is a zero energy eigenstate of
the Hamiltonian for the quantum electromagnetic field on a
globally hyperbolic manifold . On a globally hyperbolic manifold
 we can choose coordinates \cite{global} such that the metric takes
the form (we use Greek letters for space-time indices and Latin
letters for spatial indices)
\begin{equation}
ds^{2}=g_{00}dtdt-g_{jk}dx^{j}dx^{k} .\end{equation} Then, the
canonical momentum in the $A^{0}=0$ gauge is

\begin{equation}
\Pi^{j}=g^{00}\sqrt{-g}g^{jk}\partial_{0}A_{k}.
\end{equation}
The Hamiltonian is
\begin{equation}
H=\frac{1}{2}\int\frac{g_{00}}{\sqrt{-g}}g_{jk}\Pi^{j}\Pi^{k}+\frac{1}{4}\int\sqrt{-g}g^{jk}g^{ln}F_{jl}F_{kn}.
\end{equation}
In the quantized model
\begin{equation}
\Pi^{j}=-i\hbar\frac{\delta}{\delta A_{j}}.
\end{equation}
We prove that on a globally hyperbolic manifold in the metric (23)
$H\psi_{CS}=0$. For this purpose we  have to show
\begin{equation}\begin{array}{l}
\frac{1}{2}\int
g_{00}(-g)^{-\frac{1}{2}}g_{jk}\Pi^{j}\Pi^{k}\psi_{CS}=- \int
g_{00}(-g)^{-\frac{1}{2}}g_{jk}\epsilon^{jrl}\epsilon^{kmn}F_{rl}F_{mn}\psi_{CS}\cr
=\frac{1}{4}\int
\sqrt{-g}g^{jl}g^{rm}F_{jr}F_{lm}\psi_{CS}\end{array}
\end{equation}
For the proof we have calculated explicitly the elements $g^{nl}$
in terms of $g_{jk}$. We have $-g=g_{00}\det(g_{rn})$.
$g^{ln}=(\det g_{jk})^{-1}P_{2}(g_{rn})$ where $P_{2}$ is a
homogeneous polynomial of the second order in $g_{rn}$. We can see
that the second line in eq.(27) is of the form
$\sqrt{g_{00}}\det(g_{jk})^{-\frac{1}{2}}g_{rn}FF$ whereas the
third line takes the form

$\sqrt{g_{00}}\det(g_{jk})^{-\frac{1}{2}}P_{2}(g_{rl})P_{2}(g_{mn})\det(g_{jk})^{-1}FF$.
The  determinant is a homogeneous trilinear form in $g_{jk}$. We
have  shown   that the four-linear form
$P_{2}(g_{jk})P_{2}(g_{mn})FF$ divided by a trilinear form
$\det(-g_{jk})$  on the rhs of eq.(27) is the same as the linear
form $g_{jk}FF$ on the lhs. This is rather arduous but quite
elementary calculation.

The calculations are simple in the case of the  orthogonal metric
\begin{equation}
ds^{2}=a_{0}^{2}dt^{2}-a_{1}^{2}dx_{1}^{2}-
a_{2}^{2}dx_{2}^{2}-a_{3}^{2}dx_{3}^{2}.
\end{equation}
 In this metric on the lhs of eq.(27) \begin{equation}\begin{array}{l} \int
g_{00}(-g)^{-\frac{1}{2}}g_{jk}\Pi^{j}\Pi^{k}\psi_{CS}=-\int
a_{0}a_{1}^{-1}a_{2}^{-1}a_{3}^{-1}a_{j}^{2}\epsilon^{jrl}
F_{rl}\epsilon^{jmn} F_{mn}\psi_{CS}
\end{array}
\end{equation}
whereas on the rhs we have
 \begin{equation}\begin{array}{l}\frac{1}{2}\int
\sqrt{-g}g^{jl}g^{rm}F_{jr}F_{lm}\psi_{CS} \cr=\int
(a_{0}a_{3}a_{1}^{-1}a_{2}^{-1}F_{12}^{2}+a_{0}a_{2}a_{1}^{-1}a_{3}^{-1}F_{13}^{2}
+a_{0}a_{1}a_{2}^{-1}a_{3}^{-1}F_{23 }^{2})\psi_{CS}.\end{array}
\end{equation}
It is easy to see that (30) is equal to (29).

We write down the Schr\"odinger equation and the stochastic
equations in the case of isotropic orthogonal coordinates
$a_{1}=a_{2}=a_{3}$.
 In such a case the
equation for $\chi$ ( the counterpart of eq.(10)) reads
\begin{equation}
\partial_{t}\chi=\int d{\bf x} \Big(\frac{i\hbar}{2}a_{0}a_{1}^{-1}\frac{\delta^{2}}{\delta
{\bf A}({\bf x})^{2}}\pm
ia_{0}a_{1}^{-1}\epsilon_{jkl}\partial_{k}A_{l}\frac{\delta}{\delta
A_{j}({\bf x})}\Big)\chi.
\end{equation}
The stochastic equation will be a perturbation of the self-duality
equation
\begin{equation}\begin{array}{l}
F^{(\pm)}_{\mu\nu}=\pm\frac{i}{2}\tilde{\epsilon}_{\mu\nu\alpha\beta}F^{\pm\alpha\beta}=
\pm\frac{i}{2}\epsilon_{\mu\nu\alpha\beta}\sqrt{-g}F^{\pm\alpha\beta}.
\end{array}\end{equation} The stochastic equation corresponding to the
$\psi_{CS}$ state (in the time-independent isotropic orthogonal
metric) reads
\begin{equation}\begin{array}{l}
dA_{j}(s)=\pm ia_{0}a_{1}^{-1}\epsilon_{jkl}\partial_{k}A_{l}ds+
\sqrt{\frac{i\hbar
a_{0}}{a_{1}}}dB_{j}(s),\end{array}\end{equation} where the
covariance of the Brownian motion is
\begin{equation}
E[B_{j}(t,{\bf x})B_{k}(s,{\bf
y})]=min(t,s)(\delta_{jk}-\partial_{j}\partial_{k}\triangle^{-1})
\delta({\bf x}-{\bf y})\end{equation}
 The projector
$\delta_{jk}-\partial_{j}\partial_{k}\triangle^{-1}$ ensures the
fulfillment of the Gauss law \cite{habaepjplus} for the solution
of eq.(33)
$\nabla_{j}F^{0j}=(-g)^{-\frac{1}{2}}\partial_{j}(\sqrt{-g}F^{0j})=\partial_{j}\Pi^{j}=0$
in the sense that $\partial^{x}_{j}\sqrt{-g}E[F^{0j}(t,{\bf
x})F_{0k}(s,{\bf y})]=0$ .

The $\psi_{CS} $ state is not invariant under a spatial reflection
${\bf x}\rightarrow -{\bf x}$. We can built states invariant under
this transformation as a superposition
$\psi_{CS}^{(+)}\chi^{(+)}+\psi_{CS}^{(-)}\chi^{(-)}$, where $\pm$
refers to the $\pm$ signs in the CS state (22). Then, the time
evolution is generated separately for $(+)$ and $(-)$ states with
the corresponding stochastic processes differing in sign of the
drift in eq.(33).  The Brownian motion in stochastic equations is
defined for $t\geq 0$. So far in our discussion $\psi_{t}$ has
been considered only for a non-negative time. The time evolution
for $-t\geq 0$ is defined as $\psi_{t}=\overline{\psi_{-t}}$
ensuring the invariance under the time reflection in quantum
mechanics (see the discussion in \cite{habaentropy}).

The Chern-Simons wave function $\psi_{CS}$ is not square
integrable. We need $\chi$ in eq.(4) decaying sufficiently fast in
order to make $\psi$ square integrable. It comes out that  $\psi $
may preserve some crucial properties of $\psi_{CS}$ after a
multiplication by $\chi$ (treated as a regularization). As an
example we consider QED on the Minkowski space-time with
\begin{equation}
\chi=\exp(-\frac{\gamma}{2}({\bf A},\omega{\bf A})),
\end{equation}where (,) denotes the $L^{2}$ scalar product, $\omega=\sqrt{-\triangle}$ and
$\partial_{j}A_{j}=0$. The solution of the stochastic equation
(33) ($a_{0}=a_{1}=1$) in the momentum space reads
\begin{equation} {\bf A}_{t}({\bf A})=O_{t}{\bf A}
+\sqrt{i\hbar}\int_{0}^{t}O_{t-s}d{\bf B}(s),
\end{equation}
where the matrix $O_{t}$ has the matrix elements ($p=\vert{\bf
p}\vert$)
\begin{equation}
O(t)_{jk}=(\delta_{jk}-p_{j}p_{k}p^{-2})\cos(pt)-\epsilon_{jkl}p_{l}p^{-1}\sin(pt).
\end{equation}Now,
\begin{equation}
\chi_{t}({\bf A})=\psi_{CS}E[\chi({\bf A}_{t}({\bf A}))].
\end{equation} It can be checked that
\begin{displaymath}
(O_{t}{\bf A},\omega O_{t}{\bf A})=({\bf A},\omega {\bf A}).
\end{displaymath}
Hence,if we insert the solution (36) in eqs.(35) and (38) then we
obtain
\begin{equation}\begin{array}{l}
({\bf A}_{t}({\bf A}),\omega{\bf A}_{t}({\bf A}))=({\bf A},\omega
{\bf A})+\sqrt{i\hbar}({\bf A},\omega\int_{0}^{t}O_{-s}d{\bf
B}_{s})\cr+i\hbar(\int_{0}^{t}O_{-s^{\prime}}d{\bf
B}_{s^{\prime}}, \omega\int_{0}^{t}O_{-s}d{\bf B}_{s}).\end{array}
\end{equation}
We can perform the Gaussian ${\bf B}$ integral in eq.(38). The
result is of the form $\exp\Big(-\frac{1}{2}({\bf A},(\gamma\omega
+K_{t}(\hbar\gamma)){\bf A}\Big)$, where the operator
$K_{t}(\hbar\gamma)\rightarrow 0$ when $\hbar\gamma\rightarrow 0$.
 We calculate the
correlation functions (the integral over ${\bf A}$ is well-defined
for $\gamma\hbar>1$) first at $t=0$ in the state $\psi_{CS}\chi$.
We obtain
\begin{equation}
\begin{array}{l}(\psi,A_{j}({\bf p})A_{k}({\bf
p}^{\prime})\psi)=\hbar\delta({\bf p}+{\bf p}^{\prime})(
\frac{\hbar\gamma}{\hbar^{2}\gamma^{2}-1}(\delta_{jk}-p_{j}p_{k}p^{-2})
+\frac{i}{\hbar^{2}\gamma^{2}-1}\epsilon_{jrk}p^{r}p^{-2}).
\end{array}\end{equation} Eq.(40) is singular at
$\gamma\hbar=1$. It makes sense at $\hbar\gamma=0$. We may treat
the rhs of eq.(40) for $\gamma\hbar<1$ as an analytic continuation
in the complex $\gamma$ plane of the integral on the lhs of
eq.(40). A definition by an analytic continuation is necessary
even for the ordinary Feynman oscillatory integral with the
Chern-Simons Lagrangian \cite{witten1}\cite{witten2}.

Under the time evolution $U_{t}$
\begin{equation}
(U_{t}\psi,A_{j}({\bf p})A_{k}({\bf p}^{\prime})U_{t}\psi)=
(\psi,A_{j}({\bf p})A_{k}({\bf p}^{\prime})\psi)+{\cal
O}(\hbar\sqrt{\hbar}),
\end{equation}where the correction to eq.(40) is of order
$\sqrt{\hbar}$ as a consequence of eq.(39).  It follows that  if
we calculate the loop expectation values in the regularized CS
states for the loops $C_{1}$ and $C_{2}$
\begin{equation}\begin{array}{l}
(U_{t}\psi,\int_{C_{1}} {\bf A}d{\bf x}_{1} \int_{C_{2}} {\bf
A}d{\bf x}_{2}U_{t}\psi)= -\hbar 4\pi link(C_{1},C_{2})+{\cal
O}(\hbar\sqrt{\hbar}),
\end{array}\end{equation} where $link(C_{1},C_{2})$ denotes the Gauss linking number.
The expression of the loop integrals by the topologically
invariant linking number is of immense importance for quantum
gravity where only observables invariant under diffeomorphisms are
well-defined.

 We can repeat some
calculations for an electromagnetic field  on a manifold. We
cannot derive such detailed results as in eqs.(40)-(42) because
the Fourier transform is not applicable there. Nevertheless, let
us note that the operator
\begin{equation} (M_{g}{\bf A})_{j}=\frac{1}{2}\epsilon_{jkl}\sqrt{-g}g^{kr}g^{lm}F_{rm}
\end{equation}
is a Hermitian operator in the Hilbert space with the scalar
product
\begin{equation}
({\bf A},{\bf A}^{\prime})_{g}=\int d{\bf
x}\sqrt{-g}g^{00}g^{jk}\overline{A}_{j}A_{k}^{\prime}.
\end{equation}Let us define
\begin{equation}
\chi=\exp\Big(-\frac{\gamma}{2}({\bf A},\sqrt{M_{g}^{2}}{\bf
A})_{g}\Big).
\end{equation}
Then, with $O_{t}=\exp(itM_{g})$ eq.(33) has the solution of the
form (36) (with the noise ${\bf B}$ multiplied by
$\sqrt{\frac{a_{0}}{a_{1}}}$). Then, because $O_{t}=\exp(itM_{g})$
is unitary and commutes with $M_{g}$
\begin{equation}\begin{array}{l}
({\bf A}_{t}({\bf A}),\sqrt{M_{g}^{2}}{\bf A}_{t}({\bf
A}))_{g}=({\bf A},\sqrt{M_{g}^{2}}{\bf
A})_{g}+\sqrt{\hbar}q_{1}({\bf A},{\bf B})+\hbar q_{2}({\bf
B})\end{array}
\end{equation}where we used $(O_{t}{\bf A},\sqrt{M_{g}^{2}}O_{t}{\bf
A})_{g}=({\bf A},\sqrt{M_{g}^{2}}{\bf A})_{g}$. We denoted by
$q_{1}$ a form linear in ${\bf A}$ and in ${\bf B}$ and by $q_{2}$
a form bilinear in ${\bf B}$. If $\psi_{CS}\chi$ is to be square
integrable then the operator
$(M_{1})_{jk}=\epsilon_{jkl}\partial_{l}$ appearing in the
Chern-Simons form must be bounded by $\sqrt{M_{g}^{2}}$ or
equivalently for the squares of these operators we need
\begin{equation}
({\bf A},-\triangle {\bf A})<C ({\bf A},M_{g}^{2}{\bf A})_{g}.
\end{equation}with a certain constant $C$. A simple condition
 $a_{0}a_{1}^{-1}>C$ point-wise ensures the inequality (47).
However, this condition is not satisfied for the Schwarzschild
black-hole (in stationary isotropic coordinates $a_{0}$ can be
zero). Perhaps some more sophisticated estimates could work in
this case. We can perform the Gaussian integral over ${\bf B}$ in
eq.(38) in order to calculate $\psi_{t}$. A subsequent integration
over ${\bf A}$ if $C\gamma\hbar>1$ allows to calculate the
correlation functions (41)-(42). The resulting two-point function
has a limit $\hbar\gamma\rightarrow 0$ which can be considered as
an analytic continuation (in $\gamma$) of the regularized
correlation functions to their values in the Chern-Simons state
(22). In this sense the basic property (42) of the CS states will
be preserved.

 There are
minor changes in the equations of this section for non-Abelian
gauge theories (see \cite{habaepjplus} for the Minkowski
space-time) when considered on a general (hyperbolic) Lorentzian
manifolds. We just add an internal index $c$ so that
$A_{j}\rightarrow A_{j}^{c}$, $\Pi_{j}\rightarrow \Pi_{j}^{c}$ and
$F_{\mu\nu}\rightarrow F_{\mu\nu}^{c}$. As a prospective
application of the results of this paper to a unification of
quantum gravity with gauge interaction (see
\cite{mansouri}\cite{townsend} and the recent papers
\cite{cham}\cite{zoup} with references cited there) we consider
the gauge theory with the group $SO(1,n)$  as an internal symmetry
( $n=13$ suggested in \cite{cham} and $n=17$ in \cite{zoup} . We
choose as the Lie algebra the generators $J^{AB}$ (so
$c\rightarrow AB$ of the rotations in the $AB$ planes of the (n,1)
hyperboloid. We specify the notation to the SO(1,4) as in refs.
\cite{mansouri}\cite{townsend} . The gauge field is
\begin{equation}\omega_{\mu}=h^{a}_{\mu}J_{a5}+\omega^{ab}_{\mu}J_{ab},
\end{equation}where $a=1,2,3,4$. We may write an analog of the
Lagrangian (19)
\begin{equation}\begin{array}{l}
{\cal L}= =-\frac{1}{8}\sqrt{-g}F^{\pm
AB}_{\mu\alpha}F_{AB}^{\pm\mu\alpha}\mp\frac{i}{4}F^{*AB}_{\mu\alpha}F_{AB}^{\mu\alpha}.
\end{array}
\end{equation}
The last term in eq.(49) is the Chern-Simons term
$\partial_{\mu}K^{\mu}_{CS}$ where \begin{equation}
K_{CS}^{0}=Tr\int d{\bf x}(\omega
d\omega+\frac{2}{3}\omega\wedge\omega\wedge\omega).
\end{equation} In eq.(50) we expressed $K_{CS}^{0}$
 in terms of the differential form
$\omega=\omega^{AB}_{j}dx^{j}J_{AB}$ and the trace means the
invariant bilinear form $Tr(J^{AB}J^{CD})$ in the Lie algebra of
SO(1,4). The bilinear form has an indefinite sign for non-compact
groups. This does not lead to any difficulties in the Feynman
integral and in the renormalization theory of the non-compact
Yang-Mills \cite{don2} but encounters difficulties with an
interpretation of the energy spectrum of such models. From the
path integral (5) it follows that $\psi_{CS}$ is changing only by
a phase $Z_{t}$. By the argument applied in the case of the
electromagnetic field and non-Abelian gauge theories on the
Minkowski space-time \cite{habaepjplus} we can show that
$Z_{t}=1$. The Lagrangian (49) can be expanded in $\omega$ and
$h$. When we  insert the metric
\begin{equation}
g_{\mu\nu}=h_{\mu}^{a}h_{\nu}^{a}
\end{equation}
then the Lagrangian (49) consists
\cite{mansouri}\cite{townsend}\cite{cham} of the term quadratic in
the curvature tensor, the Einstein-Hilbert part (in the Palatini
form), the cosmological term and the total derivative (the
Chern-Simons term). A functional integration over $\omega$ gives
an effective action depending only on $h$ . In the approximation
neglecting the term quadratic in the curvature we obtain the
standard Palatini version of Einstein gravity.The Chern-Simons
state is still   the ``ground state" in the model (49). Being a
topological invariant it is an interesting state for a calculation
of loop correlation functions in the $SO(1,n)$ model of quantum
gravity. For the group $SO(1,n)$ ($n>4$) the decomposition (48) is
of the form
$\omega_{\mu}=\omega_{\mu}^{ab}J_{ab}+h_{\mu}^{a\alpha}J_{a\alpha}
+A_{\mu}^{\alpha\beta}J_{\alpha\beta}$ where the Greek letters
$\alpha,\beta\geq 5$, $J_{\alpha\beta}$ are generators of a
compact group and $A_{\mu}^{\alpha\beta}$ are the corresponding
gauge fields. Calculations of correlation functions, initiated for
the electromagnetic field in this section, can be extended by
perturbation theory in a three-linear Chern-Simons action (50) (
the three-linear interaction $\omega\wedge\omega\wedge\omega$ with
the propagator $ \partial^{-1}$ is renormalizable by power
counting in three dimensions).

Our approach applies also to the self-dual version
\cite{nieto}\cite{soo} ( a version of MacDowell-Mansouri model
\cite{mansouri}) of Jacobson-Smolin-Samuel gauge theory of gravity
\cite{jacobson}\cite{samuel} which is equivalent to the standard
gravity in Ashtekar's variables \cite{ashtekar}. In such a
formulation the expression of the connection in terms of the
tetrads results from the Lagrangian as a constraint.

\section{Discussion and
outlook} We think that the time-independent states (``ground
states") derived from the path integral can be useful  in a
construction of theories with a  Lagrangian of a geometric origin
like unified theories of all interactions (including gravity). The
exact quantization of such theories with an infinite dimensional
symmetry group requires a detailed study of all constraints. It
seems that for a calculation of expectation values of observables
invariant under this group a formulation of the path integral may
be simple. Then, by means of the formula  (6) we may find one of
the ``ground states". In gauge theories the  ``ground state" (the
Chern-Simons state) is expressed by a three-linear topological
invariant. A calculation of Wilson loop correlation functions in
the Chern-Simons states may lead to a well-defined unified gauge
theory of gravity and internal interactions respecting all
required symmetris. The Chern-Simons wave function together with
the Gaussian variable $Q$ can be used for a definition of the
stochastic process. Then, we may search for all states $\chi$
whose expectation values do not depend on time ( ``the wave
functions of the Universe"). The procedure seems to be manageable
in the case of $SO(1,n)$  gauge theory of gravity and its
 modification by  Jacobson and Smolin \cite{jacobson} and
Samuel \cite{samuel}. As follows from \cite{nieto} in this
formulation the Lagrangian takes the form (2) with Pontryagin and
Euler topological invariants  which can be expressed as a total
divergence.
 As a consequence we obtain a Chern-Simons
``ground state" as an analog of the Kodama ``ground state"
\cite{kodama} now formulated in the gauge theory of gravity. After
we find one of these ``ground states"  we can find all of them as
the $t\rightarrow \infty$ limit of the stochastic dynamics in the
spirit of Parisi-Wu stochastic quantization \cite{parisiwu}. A
study of the corresponding stochastic equations and their
consequences are now under investigation.

{\bf Data Availability Statement}: No Data associated in the
manuscript

  \end{document}